\documentclass[prb,twocolumn,showpacs,superscriptaddress]{revtex4}

\usepackage{amsmath,amssymb}
\usepackage{graphicx}

\newcommand{\void}[1]{}
\begin{document}

\title{Discontinuous conductance of bichromatically ac-gated quantum wires}

\author{Tomasz Kwapi\'nski}
\email{tomasz.kwapinski@umcs.lublin.pl}
\affiliation{Institut f\"ur Physik, Universit\"at Augsburg,
Universit\"atsstra{\ss}e 1, D-86135 Augsburg, Germany}
\affiliation{Institute of Physics, M.~Curie-Sk\l odowska University,
20-031 Lublin, Poland}

\author{Sigmund Kohler}%

\author{Peter H{\"{a}nggi}}%
\affiliation{Institut f\"ur Physik, Universit\"at Augsburg,
Universit\"atsstra{\ss}e 1, D-86135 Augsburg, Germany}

\date{\today}

\begin{abstract}
We study the electron transport through a quantum wire under the
influence of external time-dependent gate voltages. The wire is
modelled by a tight-binding Hamiltonian for which we obtain the
current from the corresponding transmission.  The numerical evaluation
of the dc current reveals that for bichromatic driving, the
conductance depends sensitively on the commensurability of the driving
frequencies.  The current even possesses a discontinuous frequency
dependence.  Moreover, we find that the conductance as a function of
the wire length oscillates with a period that depends on the ratio
between the driving frequencies.
\end{abstract}


\maketitle

\section{\label{sec10}Introduction}

Recent advancements in nanotechnology enabled the
investigation of low-dimensional systems like quantum dots,
quantum wires or few-atom systems.  Thereby,
techniques like mechanically controlled break junctions,
scanning tunneling microscopy, or angle-resolved photo-emission
electron spectroscopy provide information about the electronic
structure or transport properties of these systems. Particularly
interesting are the transport properties of low-dimensional nanoscale
conductors in the presence of time-dependent external fields for
which a wealth of effects has been predicted like, e.g.,
photon-assisted tunneling,\cite{Oost,Wiel} and electron pumping.
\cite{Sta,Kouw} External fields can be also used to control both
the dc current \cite{Leh} and the zero-frequency noise \cite{Cam,Cam2}
in mesoscopic conductors.

In undriven one-dimensional quantum wires, conductance
oscillations --- the oscillatory dependence of the conductance as a
function of the wire length --- were predicted \cite{Lang,Zeng} and
experimentally confirmed.\cite{Smit}  They emerge when the energy
levels of a quantum wire match the Fermi energy of the system and
relate to charge neutrality of the wire.\cite{Sim}  The period of
these oscillations is usually two sites for undriven wires, but may
become larger if the wire is exposed to a time-dependent field.
\cite{Kwap, Martinez2006a, Martinez2008a}

A second harmonic added to an ac field that drives a quantum wire may
significantly alter transport properties.  The main reason of the
differences is that the second harmonic may brake time-reversal
symmetry and generalized parity dynamically, leading to effects like
non-adiabatic electron pumping,\cite{Leh2,Mos,Koh2,Mah,Arra} i.e., to
a dc current even in the absence of any net bias.  However, if the ac
field acts as a gate voltage that shifts all wire levels uniformly,
the pump current must vanish.\cite{Koh2}  Similar effects have been
predicted for quantum Brownian motion in bichromatically driven
infinitely extended periodic potentials.\cite{Goy,Sen,Bor} Moreover,
two incommensurable rectangular-shaped driving forces can cause a dc
current\cite{Sav,Sav2} and lead to interaction-induced frequency
mixing.\cite{Thor} Also dynamical localization effects in quantum dots
may be affected by bichromatic driving fields, in particular when the
freuquencies of the latter are commensurable.\cite{Bas,Wang}

A well established theoretical tool for treating periodically driven
quantum system is Floquet theory, which is based on the discrete
time-translation invariance of the driving field.  Higher harmonics of
the fundamental driving frequency do not affect this invariance and,
thus, can be included straightforwardly.  As soon as the ratio between
the fundamental frequency and its higher harmonic increases, a numerical
treatment may require quite some effort.  The situation becomes even
worse, when the frequencies are no longer commensurable.  Then it is
rather desirable, even from a practical point of view, to find
analytical expressions for the behavior of the driven quantum system,
as for example in the recent works \onlinecite{Goy,Sen,Bor}.
For the description of electron transport through monochromatically
driven systems, common transport theories have been combined with
Floquet theory.\cite{Bruder1994a,Platero2004a,Koh2,Arrachea2006a}

In this paper we study the transport properties of a bichromatically
driven quantum wire, focussing on the influence of commensurability of
the driving frequencies.  In doing so, we generalize in
Sec.~\ref{sec20} recent theoretical approaches such that they allow
one to treat two ac fields with arbitrary frequency ratio.  Then we
investigate in Sec.~\ref{sec30} the influence of bichromatic driving
on the transmission and the conductance oscillations.

\section{\label{sec20}Theoretical description}
\subsection{Wire-lead model}

We consider the time-dependent wire-lead Hamiltonian $H(t) =
H_\mathrm{wire}(t) + H_\mathrm{leads} + H_\mathrm{coupl}$, where
\begin{equation}
\label{model}
H_\mathrm{wire}(t) =  \sum_{n=1}^N [\varepsilon_n + f(t)] c^+_{n}
c_{n}
  +\sum^{N-1}_{n=1} V_{n} c^+_{n} c_{n+1} + {\rm h.c.}
\end{equation}
models in a tight-binding description the wire with sites
$|n\rangle$, where neighboring sites are coupled via tunnel matrix
elements $V_n$. The on-site energies $\varepsilon_n$ are modulated
by an ac gate voltage $f(t)$.  The first and the last site of the
wire are coupled via the tunnel Hamiltonian
\begin{equation}
H_\mathrm{coupl}
= \sum_{\vec kL} V_{\vec kL} c^+_{\vec kL}c_1 +
  \sum_{\vec kR} V_{\vec kR} c^+_{\vec kR} c_N
\label{h2}
\end{equation}
to leads at the left ($L$) and at the right ($R$) end.  Within a
wide-band approximation, we assume that the spectral density of
the wire-lead coupling, $\Gamma_\alpha(\varepsilon) \equiv
2\pi\sum_{\vec k} |V_{\vec k}|^2
\delta(\varepsilon-\varepsilon_\alpha) =\Gamma_\alpha$, is energy
independent.
The leads are modelled as ideal Fermi gases by the Hamiltonian
\begin{equation}
H_\mathrm{leads}
= \sum_{\alpha=L,R} \sum_{\vec k}
  \varepsilon_{\vec k\alpha} c^+_{\vec k\alpha} c_{\vec k\alpha} .
\label{h1}
\end{equation}
The operators $c_{\vec k\alpha}$, $c_n$, and their Hermitian
adjoints are the usual annihilation and creation operators for
an electron in lead $\alpha = L, R$ with wave vector $\vec k$ and
for an electron at the $n$th wire site, respectively.
We focus on experiments like those described in
Refs.~\onlinecite{Zeng,Kra,Kra2}, in which electron-electron
interactions can be neglected.  For photon-assisted transport, this
is, e.g., the case when the photon energy is much smaller than both
the typical splitting between single-particle energies of the wire and
the Coulomb interaction.\cite{Zhao}

\subsection{Time-dependent scattering theory}

Following the scattering approach of Ref.~\onlinecite{Koh2}, we
derive from the time-dependent wire-lead Hamiltonian $H(t)$ the
Heisenberg equations of motion for the creation operators.  The
wire part of their solution can be expressed in terms of a
retarded Green function $G(t,t')$ which, owing to the
time-dependence of the Hamiltonian, depends explicitly on both times
and fulfills the equation of motion
\begin{equation}
\label{eom:G}
\Big[ i\hbar\frac{d}{dt} - H_0 - f(t) +\frac{i}{2}\Gamma \Big]G(t,t')
= \delta(t-t').
\end{equation}
The Hamiltonian $H_0 = H_\mathrm{wire}-f(t)\mathbf{1}_\mathrm{wire}$
describes the wire in the absence of the time-dependent gating, while
$\Gamma = \Gamma_L |1\rangle\langle 1| + \Gamma_R |N\rangle\langle N|$
is a self-energy stemming from the coupling of the wire to the leads.

The solution of Eq.~\eqref{eom:G} can be traced back to the solution
of the time-independent problem for which the Green function $g$ obeys
the equation of motion
\begin{equation}
\label{eom:g}
\Big[ i\hbar\frac{d}{dt} - H_0  +\frac{i}{2}\Gamma \Big]g(t-t')
= \delta(t-t').
\end{equation}
Then it is straightforward to show that the Green functions $G$ and
$g$ are related according to
\begin{equation}
\label{Gg}
G(t,t') = e^{-i[F(t)-F(t')]} g(t-t').
\end{equation}
This means that the influence of the driving field is subsumed in the
phase
\begin{equation}
F(t) = \int^t ds f(s) .
\end{equation}

Defining the current through the wire as the change of the
electron number in the left lead, $I=-edn_L(t)/dt$, we obtain for
its expectation value the time-dependent Landauer-like expression
\begin{equation}
\label{I(t)}
I(t) = \frac{e}{h}\int d\varepsilon \big[
       T_{RL}(t,\varepsilon) f_L(\varepsilon)
      -T_{RL}(t,\varepsilon) f_L(\varepsilon) \big]
       -\dot q_L(t)
\end{equation}
with the time-dependent transmission
\begin{equation}
\label{T_LR}
T_{LR}(t,\varepsilon) = \Gamma_L\Gamma_R |G_{1,N}(t,\varepsilon)|^2
\end{equation}
and $T_{LR}$ defined accordingly.  The function
\begin{eqnarray}
G(t,\varepsilon)
&=& \int d\tau\, e^{i\varepsilon\tau} G(t,t-\tau)
\\
&=& e^{-i F(t)} \int d\tau\, e^{i\varepsilon\tau +iF(t-\tau)} g(\tau)
\label{G(t,e)}
\end{eqnarray}
is the Fourier transformed of the Green function with respect to the
time difference $\tau = t-t'$.  Note that $G$ still depends explicitly
on the final time $t$.  The displacement current $\dot q_L(t)$
describes charges oscillating between the left lead and the wire and
does not contribute to the dc current considered below.  Thus it will
not be considered henceforth.

In order to evaluate the $\tau$-integration in Eq.~\eqref{G(t,e)}, we
have to specify the time-dependent gate voltage $f(t)$. We assume that
it consists of $K$ harmonics, such that
\begin{eqnarray}
f(t) &=& \sum_{k=1}^K \Delta_k \cos(\omega_k t) , \label{e0(t)}
\\
F(t) &=& \sum_{k=1}^K \frac{\Delta_k}{\omega_k}\sin(\omega_k t) ,
\end{eqnarray}
where $\omega_k$ and $\Delta_k$ are the frequency and the amplitude of
the $k$th harmonic of the ac field.
Then the Green function \eqref{G(t,e)} reads
\begin{equation}
\begin{split}
G(t,\varepsilon)
=&  \sum_{m_1,\cdots,m_K} J_{m_1}(\Delta_1/\omega_1)
    \cdots J_{m_K}(\Delta_K/\omega_K)
\\
& \times \exp\big({i\sum_k m_k\omega_k t}\big)
         g\big(\varepsilon-\sum_k m_k\omega_k \big) .
\end{split}
\end{equation}
The Bessel functions $J_m$ stem from a Fourier decomposition of the
exponential in Eq.~\eqref{Gg}, while the evaluation of the $\tau$
integral provided the Fourier transformed of the Green function,
$g(\epsilon)$.  Note that we have ignored a time-dependent but
nevertheless irrelevant phase factor.

Assuming moreover symmetric wire-lead coupling, i.e., $\Gamma_L =
\Gamma_R = \Gamma$, the time-dependent transmission \eqref{T_LR}
becomes
\begin{align}
\label{T(t)}
T_{LR}(t,\varepsilon)
={}& \Gamma^2 \sum_{m_1,\cdots,m_K}
   J_{m_1}(\Delta_1/\omega_1) \cdots J_{m_K}(\Delta_K/\omega_K)
\nonumber \\ & \times
   \sum_{m_1',\cdots,m'_K}
   J_{m_1'}(\Delta_1/\omega_1)\cdots J_{m_K'}(\Delta_K/\omega_K)
\nonumber \\ & \times
   g_{1N}\big(\varepsilon-\sum_k m_k\omega_k \big)
   g_{1N}^*\big(\varepsilon-\sum_k m_k'\omega_k \big)
\nonumber \\ & \times
   \exp\big({i\sum_k (m_k - m_k') \omega_k t}\big)
\\
={} & T_{RL}(t,\varepsilon) .
\end{align}
The relation in the last line means that from the right lead to the
left lead has the same probability as the reversed process.  This is
particular for ac gating and does not hold for general
driving.\cite{Koh2}  Here it stems from the fact that for the present
model with time-dependent gating, the time-dependent transmission can
be expressed in terms of the Green function $g$ for the
time-independent problem.

The dc current, being the main quantity of interest, is then obtained
from Eq.~\eqref{I(t)} and reads
\begin{equation}
\label{I0}
I_0 = \frac{e}{h}\int d\varepsilon
      \big[ f_L(\varepsilon) -f_L(\varepsilon) \big]T(\varepsilon) ,
\end{equation}
where $T(\varepsilon)$ denotes the time-average of the transmission
$T_{LR}(t,\varepsilon)$, and its computation from Eq.~\eqref{T(t)}
seems straightforward.  This is however not the case, because the
averaging procedure depends crucially on the commensurability of the
frequencies $\omega_k$.  If the gate voltage $f(t)$ consists of many
spectral components, the actual computation can be quite cumbersome.
However, we below restrict ourselves to two frequencies, so that it is
sufficient to address two cases: all frequencies being commensurable or
all being incommensurable.

For incommensurable frequencies, the phase factor in the last line of
the time-dependent transmission~\eqref{T(t)} vanishes if and only if
$m_k = m_k'$ for each mode $k$.  Consequently, we obtain the average
transmission
\begin{eqnarray}
\label{T incom2}
T(\varepsilon)
&=& \sum_{m_1,\cdots,m_K} \Gamma^2
   J_{m_1}^2(\Delta_1/\omega_1) \cdots J_{m_K}^2(\Delta_K/\omega_K)
\nonumber \\ && \times
   \Big|g_{1N}\big(\varepsilon-\sum_k m_k\omega_k \big)\Big|^2 .
\end{eqnarray}
The Green function of the undriven wire, $g(\epsilon)$, depends on the
specific wire model and will be determined below.

If the frequencies of the external fields are commensurable,
time-averaging unfortunately does not result in a less involved
expression for the transmission.  Therefore we provide the explicit
result only for the case of two frequencies with ratio
$\omega_1/\omega_2 = m/m'$, where $m$ and $m'$ are integers. Then we
obtain
\begin{equation}
\begin{split}
T(\varepsilon)
= \sum_{{<}m_1,m_2,m_1',m_2'{>}}
& \Gamma^2
   J_{m_1}(\Delta_1/\omega_1) J_{m_2}(\Delta_2/\omega_2)
\\& \times J_{m_1'}(\Delta_1/\omega_1) J_{m_2'}(\Delta_2/\omega_2)
\\& \times g_{1N}(\varepsilon-m_1\omega_1-m_2\omega_2)
\\& \times g_{1N}^*(\varepsilon-m_1'\omega_1-m_2'\omega_2) ,
\end{split}
\label{T com}
\end{equation}
where the angular brackets restrict the sum to
those values of $m_1,m_2,m_3,m_4$ that comply with the condition
$(m_1-m_1')\omega_1 = -(m_2-m_2')\omega_2$.  Note that for
commensurable frequencies, this condition can also be fulfilled for
$m_k\neq m_k'$. Since for incommensurable frequencies the
restriction to the sum is fulfilled only for $m_1=m_1'$ and
$m_2=m_2'$, expression \eqref{T com} incidentally holds for both
cases.

The transmissions~\eqref{T incom2} and \eqref{T com} represent the
main analytical results of this work.  In the case of commensurable
frequencies, the driving is periodic and, thus, can be treated within
Floquet scattering theory.\cite{Cam,Koh2,Kwap}  Nevertheless, the
explicit evaluation of the transmission can be quite cumbersome.
The case of incommensurable frequencies generally even requires a
``two-color Floquet theory'' which can be numerically demanding.
Only the restriction to models with uniform gating allowed us to obtain
the present compact expressions for the transmission.

\section{wire with equal on-site energies}
\label{sec30}

In order to present explicit results, we have to specify
the wire model.  A $N$-site model that allows even further
analytical treatment is one with equal onsite energies
$\varepsilon_n = \varepsilon_0$ for all sites $|n\rangle$,
$n=1,\ldots,N$ and equal couplings $V_n=V$. For such a wire, the
relevant element of the Green function in the absence of driving
reads
\begin{equation}
g_{1N}(\varepsilon) = \frac{(-1)^{N-1}}{V u_N(x) + i\Gamma
u_{N-1}(x) - \Gamma^2 u_{N-2}(x)/4V} , \label{chebysch}
\end{equation}
where $x = (\varepsilon-\varepsilon_0)/2V$, while $u_n$ denotes
the $n$th Chebyschev polynomial of the second kind.  For a
derivation, we refer the reader to Ref.~\onlinecite{Kwap2}.
This relation together with Eqs.~\eqref{T incom2}
and \eqref{T com} allows one to obtain explicit expressions for the
transmission of the ac-gated wire~\eqref{model}.

In our calculations we use the effective wire-lead coupling strength
$\Gamma$ as energy unit.  Then for $\Gamma = 0.1\mathrm{eV}$, the
current unit becomes $2e\Gamma/\hbar \simeq 50$ $\mu$A. Moreover, we
measure level energies with respect to the Fermi energy, which implies
$E_F=0$.  Henceforth, we present results for a bichromatic driving field.

\subsection{Discontinuous transmission and corresponding current}
\label{sec32}

\begin{figure}[t]
\begin{center}
\includegraphics[width=0.9\columnwidth]{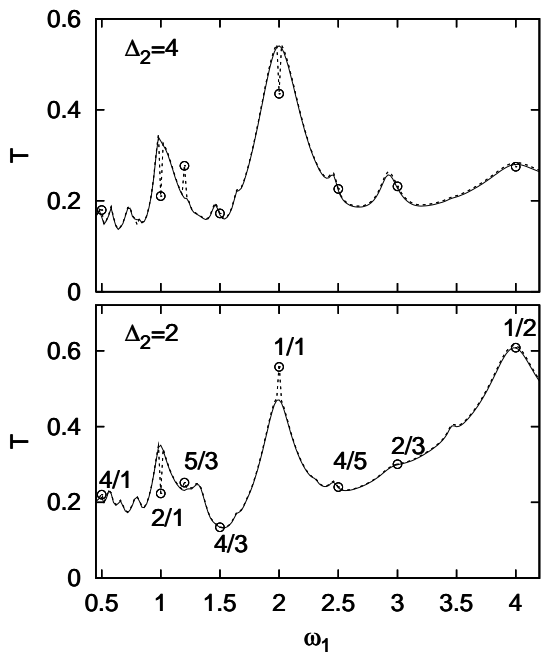}
\end{center}
\caption{Transmission for a wire with $N=5$ sites with equal
onsite energies $\varepsilon_0=0$ as a function of the frequency
$\omega_1$ for the driving amplitudes $\Delta_2=4$ (a)
and $\Delta_2=2$ (b).  The tunnel coupling between two
neighboring wire sites is $V=4$, while the other driving
parameters read $\omega_2=2$, $\Delta_1=4$. Solid lines are
obtained from expression~\ref{T incom2} which is valid for
incommensurable frequencies only, while the dashed lines and the
circles are obtained from Eq.~\eqref{T com} and, thus, are valid
for all driving frequencies $\omega_1$, $\omega_2$.  The fractions
in panel (b) mark selected frequency ratios
$\omega_2/\omega_1$.} \label{fig02}
\end{figure}%
In order to reveal the role of commensurability of the driving
frequencies, we investigate the transmission of a wire with onsite
energies at the Fermi energy level. Figure~\ref{fig02} shows the
result as a function of the frequency of the first external field,
$\omega_1$, while the frequency of the second field is kept at a
constant value.  As a most prominent feature, the transmission turns
out to be a discontinuous function, which assumes exceptional values
whenever $\omega_2/\omega_1$ is rational, i.e., when the driving
frequencies are commensurable.  A clearly visible effect is found
for ratios that can be expressed by small natural numbers.  In
particular, for the ratios $1/1$ and $2/1$, the deviation from the
transmission in the vicinity of the resonance is of the order
of 20\%.  For ``less rational'' numbers like e.g.\ $4/3$ or $2/3$,
we still find a small discontinuity which, however, is hardly
visible on the scale chosen.  The values can depend sensitively on
details like driving amplitudes and wire parameters.  In the
present case, it is even such that the deviation at the prime
resonance $1/1$ turns with increasing amplitude $\Delta_2$ from a
remarkable negative value to a positive value; cf.\ panels (a) and
(b) of Fig.~\ref{fig02}.  Moreover, the $2/1$ resonance,
albeit ``very rational'', does here not lead to a significant
discontinuity.  A similar behavior has been observed for the
transport in classical ratchet devices.\cite{Sav2} It is worth
noting that the transmission for incommensurable frequencies
[Eq.~\eqref{T incom2} and solid lines in Fig.~\ref{fig02}]
possesses local maxima at the resonances $1/2$, $1/1$, and $2/1$.

\begin{figure}[tb]
\begin{center}
\includegraphics[width=0.9\columnwidth]{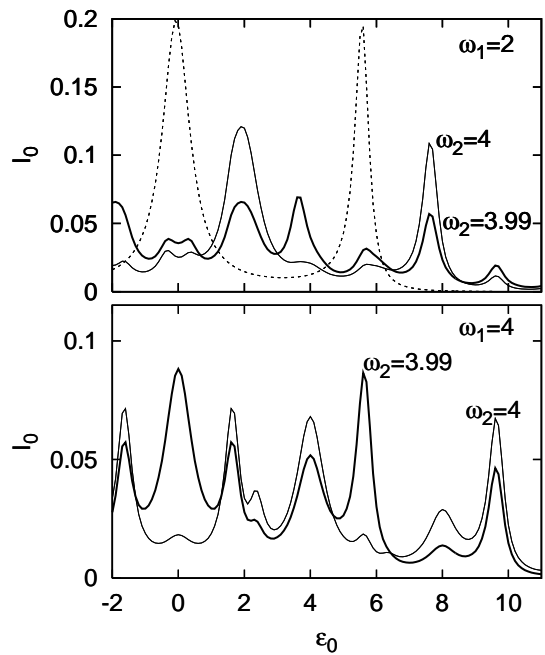}
\end{center}
\caption{Average current for the chemical potentials
$\mu_L=-\mu_R=0.1$ as a function of the equal onsite energies
$\varepsilon_0$ for a wire with $N=3$ sites and tunneling matrix
elements $V=4$.  The first driving field is kept constant
[$\Delta_1=4$ with frequency $\omega_1=2$ (a) and
$\omega_1=4$ (b)].  The second field has driving
amplitude $\Delta_2=4$, while its frequency is $\omega_2 = 4$
(thick solid lines) and $\omega_2=3.99$ (thin solid lines),
respectively.  The dashed line in panel (a) marks the
undriven case which corresponds to vanishing driving amplitudes,
i.e.\ $\Delta_1 = \Delta_2 =0$.} \label{fig03}
\end{figure}
In order to investigate further the role of commensurability, we focus
on the 2/1 resonance for $\omega_1=2$ and the 1/1 resonance for
$\omega_1=4$.  In doing so, we compare the current for the
commensurable case $\omega_2=4$ with the one found for the
``less commensurable'' $\omega_2=3.99$.  The current for a wire of
length $N=3$ as a function of the onsite energy $\varepsilon_0$ is
shown in Fig.~\ref{fig03}.
A common feature of both cases is the emergence of side peaks: The
undriven wire possesses the eigenenergies $E_0=\varepsilon_0$ and
$E_\pm=\varepsilon_0\pm5.6$.  Whenever these eigenenergies lie
within the voltage window at the Fermi energy $E_F = 0$, i.e.\ for
$\varepsilon_0=0$ and for $\varepsilon_0=\mp 5.6$, the current
assumes a maximum in compliance with the Breit-Wigner formula; see
Fig.~1 of Ref.~\onlinecite{Kwap2}.

In the presence of ac gating, we find side peaks shifted from
the original peaks by multiples of the driving frequencies $\omega_1$
and $\omega_2$.
Overlapping peaks may lead to destructive interference.  This is,
e.g., the case for the data shown in
Fig.~\ref{fig03}(a) in the vicinity of $\varepsilon_0=0$, where the
side peaks at $\varepsilon_0 \mp\omega_1 \mp\omega_2$ for
$\varepsilon_0=\pm 5.6$ compensate the fundamental peak at
$\varepsilon_0=0$, such that the total current is characterized by
the local minimum.
The proper location of the side peaks is most clearly visible for
$\omega_1=4$ and $\omega_2\approx 4$ [Fig.~\ref{fig03}(a)], i.e., when
both frequencies are practically indistinguishable.  Then we observe
side peaks at $\varepsilon_0 = 4$, $8$ (main peak at $\varepsilon_0 =
0$), at $\varepsilon_0 = 1.6$, $9.6$ (main peak at $\varepsilon_0 =
5.6$), and at $\varepsilon_0 = 2.4$ (main peak at
$\varepsilon_0=-5.6$).
Let us emphasize that the location of the peaks does not allow us
to divide the commensurable case from the incommensurable case, because
the resonance width $\Gamma=1$ is much larger than the difference
between the two values for $\omega_2$ which reads $4-3.99 = 0.01$.

Despite the practically identical location of the peaks, the peak
heights are significantly different for the two cases. This
corresponds to the discontinuity observed already in Fig.~\ref{fig02}.
The difference can even be up to a factor 5, as is the case for
$\omega_1=4$ [Fig.~\ref{fig03}(b)] for the zeroth-order resonances at
$\varepsilon_0 = 0$ and $\varepsilon_0 = 5.6$.  This means that in a
possible experiment, the current may be controlled via a slight shift
of one of the driving frequencies, provided that one tunes the wire
parameters such that the current in the absence of the driving assumes
a large value.

\subsection{\label{sec33} Conductance oscillations}

The conductance $G$ of a tight-binding system with onsite energies
at the Fermi level obeys oscillations as a function of the wire
length.  Typically these oscillations have period 2 (even-odd
oscillations),\cite{Lang,Zeng,Smit,Sim} but also larger periods have
been predicted for both static\cite{Per,Thy,Kwap2,Mol} and
monochromatically driven\cite{Kwap} wires.  We here investigate
the corresponding behavior for bichromatic gating.

The oscillation period for the driven case can be estimated along
the lines followed in Ref.~\onlinecite{Kwap}: For a static
wire, the emergence of conductance oscillation with period $M$
means that a wires with $N$ sites possesses the same conductance as a
wire with $N+M$ sites, i.e.
\begin{equation}
\label{co-condition}
g_{1,N}(\varepsilon) = g_{1,N+M}(\varepsilon).
\end{equation}
After some algebra, one obtains\cite{Kwap2} $\cos(m\pi/M) =
(\varepsilon-\varepsilon_0)/2 V$, where $m = 0,1,\ldots,M-1$. The
energy difference on the right-hand side of this equation marks
the distance of the onsite energies $\varepsilon_0$ to the Fermi
energy of the leads attached.  In the presence of ac fields,
onsite energies acquire shifts by multiples of the driving field
quanta.  Then the oscillation period is determined by the
condition
\begin{equation}
\cos\left({\frac{m\pi}{ M}}\right)
={\frac{\varepsilon-\varepsilon_0 + k_1 \omega_1 + k_2 \omega_2}{2
V}} , \label{oscyl}
\end{equation}
where again $m=0,1,\ldots,M-1$, while $k_1$ and $k_2$ are integer
numbers.

\begin{figure}[tb]
\begin{center}
\includegraphics[width=0.85\columnwidth]{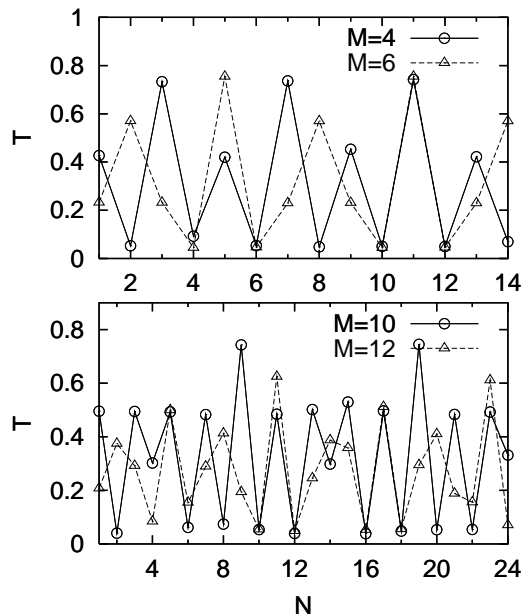}
\end{center}
\caption{Transmission as a function of the wire length $N$ for
$\Delta_1=\Delta_2=6$,  $\varepsilon_0=0$, $V=4$. (a) the
driving frequencies are $\omega_1=\sqrt 2 V$, $\omega_2=2V$
(circles) and $\omega_1=V$, $\omega_2=2V$ (triangles) correspond
to the oscillation periods $M=4$ and $M=6$, respectively.
(b) $\omega_1=2V \cos(\pi/5)$, $\omega_2=2V$ (circles) and
$\omega_1=V$, $\omega_2=\sqrt 2 V$ (triangles) corresponding to
$M=10$ and $M=12$, respectively. The lines serve as a guide to the
eye.} \label{fig04}
\end{figure}%
In order to corroborate this condition, we have computed the
transmission [which relates to the conductance
according to $G=(2e^2/h)T$] for bichromatic driving.  The results
for various combinations of the driving frequencies are shown in
Fig.~\ref{fig04}. For all parameters used in the numerical
evaluation of the transmission \eqref{T com}, we observe an
oscillation period $M>2$ in compliance with Eq.~\eqref{oscyl}.
Conductance oscillations with period $M=6$ have also been found in
Ref.~\onlinecite{Kwap} for wires under the influence of a
monochromatic ac field.  There the oscillation period is also
determined by Eq.~\eqref{oscyl} for $k_2=0$.  We like to emphasize
that in the absence of the driving, by contrast, the oscillation
period is $M=2$, i.e., we find the usual even-odd oscillations.

\begin{figure}[t]
\begin{center}
\includegraphics[width=0.85\columnwidth]{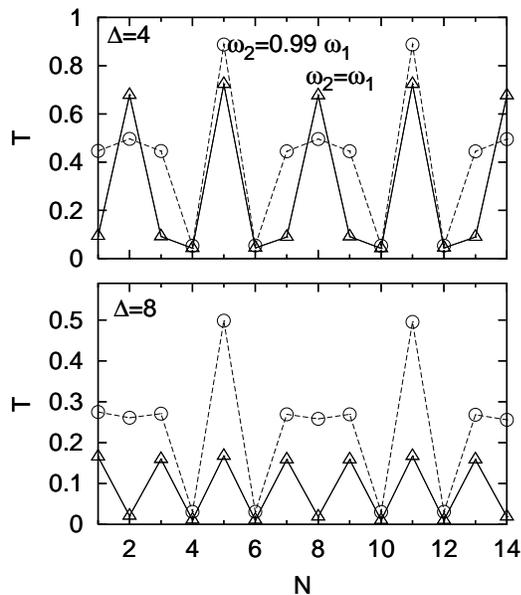}
\end{center}
\caption{Transmission as a function of the wire length $N$ for
commensurable frequencies ($\omega_1=\omega_2 =4$, triangles) and
incommensurable frequencies ($\omega_2=0.99\omega_1$,
$\omega_1=4$, circles) and the driving amplitudes
$\Delta_1=\Delta_2=\Delta=4$ (a) and
$\Delta_1=\Delta_2=\Delta=8$ (b). The other parameters
are $\varepsilon_0=0$, $V=4$. The lines are a guide to the eye. }
\label{fig05}
\end{figure}%
A remaining intriguing question is whether for bichromatic
driving, the commensurability of the frequencies influences the
conductance oscillations.  Therefore, we compare for $\omega_1 =
V$  the commensurable $\omega_2 = \omega_1$ (which in fact represents
monochromatic driving) with the ``practically incommensurable''
$\omega_2 = 0.99\omega_1$. For both cases, Eq.~\eqref{oscyl}
predicts conductance oscillations with a period of approximately
$M=6$. Figure \ref{fig05} depicts the resulting transmission as a
function of the wire length.  Note, that the oscillation period
for the static conductor is $M=2$, while it becomes $M=6$ in the
presence of the driving.
Generally, the variation of the conductance within one oscillation
period is larger for incommensurate driving frequencies.  For the
driving parameters used in Fig.~\ref{fig05}, it is even such that for
the commensurate $\omega_1=\omega_2$, the oscillations have almost
period $M=3$ or $M=2$ (solid lines) with a small variation of every
second peak, such that strictly speaking, the period is still $M=6$.
Moreover, we again observe one significant difference between the
commensurate and the incommensurate case: For given driving
amplitudes, the conductance may change by almost one order of magnitude
upon shifting the second frequency from $\omega_2=\omega_1$ to
$\omega_2 = 0.99\omega_1$.  This again allows one in an experiment to
switch the current by changing one driving frequency by only $1\%$.

Note also that when the driving amplitude becomes smaller, the
situation tends towards the undriven limit for which the oscillation
period is $M=2$.  The transition is such that the conductance within
the $M=6$ period changes until eventually the oscillation period
becomes practically $M=2$.

\section{\label{sec40}Conclusions}

Using a Green's function technique, we have derived a Landauer-like
formula for the current through a polychromatically ac-gated quantum
wire described by a tight-binding model.  If the gating acts uniformly
on all wire sites, we have provided an expression for the
time-dependent transmission that depends on the transmission in the
absence of the driving and some time-dependent prefactors.  The
time-average of the transmission for ac gating determines the dc
current being the central experimentally accessable quantity.  In
order to perform the time-average, we had to distinguish between
commensurable and incommensurate frequencies.  We have found that the
transmission and, thus, the dc current depends sensitively on the
commensurability of the driving frequencies.  The difference is most
noticeable close to the resonances with the ``very rational''
frequency ratios $1/2$ and $1/1$, where the conductance as a function
of one driving frequency is discontinuous and may suddenly change by
up to one order of magnitude.

The commensurability of the driving frequencies leaves also its
fingerprints in the conductance oscillations: While for static
conductors, one usually observes an oscillation period of two sites
(even-odd oscillations), bichromatic driving may increase the period
significantly.  For our model, we observed oscillation periods of up
to 12 sites.  The physical reason for the enlarged period is that the
bichromatic driving influences the resonance condition that determines
the oscillations.  We have confirmed this intuitive prediction by
numerical studies.  As a remarkable influence of the commensurability,
we found that the growth of the oscillation period is more visible for
incommensurate frequencies.

In a possible experiment with STM tips that contact a quantum
wire on a vicinal surface,\cite{Kra,Kra2} the observed effect may be
used for switching the current upon slightly changing the
frequency of the ac gate voltage.  Alternatively, such an experiment
may also be performed with a molecular wire that bridges a break
junction\cite{Smit,Agr} or with a chain of coherently coupled quantum
dots.\cite{Blick, Schroer2007a}  In particular the latter type of
experiment is well suited for applying time-dependent gate voltages.

\begin{acknowledgments}
This work has been supported by the Grant No.\ N\,N202\,1468\,33 of the
Polish Ministry of Science and Higher Education and the Alexander
von Humboldt Foundation (TK).  SK and PH acknowledge support by
the DFG via SPP 1243, the excellence cluster ``Nanosystems
Initiative Munich'' (NIM), and the German-Israeli Foundation (GIF).
\end{acknowledgments}

\end{document}